\documentclass[showpacs,twocolumn,prb,superscriptaddress,notitlepage,nofootinbib]{revtex4-2}
\usepackage[dvips]{graphicx}
\usepackage{amsmath,amssymb,amsthm,mathrsfs,amsfonts,dsfont}
\usepackage{subfigure, epsfig}
\usepackage{braket}
\usepackage{bm}
\usepackage{enumerate}
\usepackage{color}
\usepackage{chemformula}
\usepackage{siunitx}
\usepackage{threeparttable}
\usepackage{booktabs}
\usepackage[normalem]{ulem}
\usepackage{graphicx}% Include figure files
\usepackage{dcolumn}% Align table columns on decimal point
\usepackage{bm}% bold math

\usepackage[plainpages=false,colorlinks=true,linkcolor=blue,citecolor=blue,anchorcolor=blue,urlcolor=blue ]{hyperref}
\begin{document}
	\title{Magnetic excitations in the topological semimetal YbMnSb$_2$}
\author{Siobhan M. Tobin}
\affiliation{Department of Physics, University of Oxford, Clarendon Laboratory, Oxford OX1 3PU, United Kingdom}
\author{Jian-Rui Soh}
\affiliation{Institute of Physics, Ecole Polytechnique Fédérale de Lausanne (EPFL), CH-1015 Lausanne, Switzerland}
\author{Hao Su}
\affiliation{School of Physical Science and Technology, ShanghaiTech University, Shanghai 201210, China}
\author{Andrea Piovano}
\affiliation{Institut Laue-Langevin, 6 rue Jules Horowitz, BP 156, 38042 Grenoble Cedex 9, France}
\author{Anne Stunault}
\affiliation{Institut Laue-Langevin, 6 rue Jules Horowitz, BP 156, 38042 Grenoble Cedex 9, France}
\author{J. Alberto Rodr\'iguez-Velamaz\'an}
\affiliation{Institut Laue-Langevin, 6 rue Jules Horowitz, BP 156, 38042 Grenoble Cedex 9, France}
\author{Yanfeng Guo}
\affiliation{School of Physical Science and Technology, ShanghaiTech University, Shanghai 201210, China}
\affiliation{ShanghaiTech Laboratory for Topological Physics, ShanghaiTech University, Shanghai 201210, China}
\author{Andrew T. Boothroyd}
\affiliation{Department of Physics, University of Oxford, Clarendon Laboratory, Oxford OX1 3PU, United Kingdom}
%\author{Author list}
\date{\today}% It is always \today, today,
%  but any date may be explicitly specified
\begin{abstract}

We report neutron scattering measurements on YbMnSb$_2$ which shed new light on the nature of the magnetic moments and their interaction with Dirac fermions. Using half-polarized neutron diffraction we measured the field-induced magnetization distribution in the paramagnetic phase and found that the magnetic moments are well localised on the Mn atoms. Using triple-axis neutron scattering we measured the magnon spectrum throughout the Brillouin zone in the antiferromagnetically ordered phase, and we determined the dominant exchange interactions from linear spin-wave theory. The analysis shows that the interlayer exchange is  five times larger than in several related compounds containing Bi instead of Sb. We argue that the coupling between the Mn local magnetic moments and the topological band states is more important in YbMnSb$_2$ than in the Bi compounds.

\end{abstract}

\maketitle
%%%%%%%%%%%%%%%%%%%%%%%%%%%%%%%%%%%%%%%%%%%%%%%%%%%%%
\section{Introduction}
Topological metals and semimetals have excellent transport qualities as a result of special geometrical properties of the quasiparticles associated with linear electronic band crossings near to the Fermi energy. Magnetism can dramatically influence topological electronic states, and offers the possibility of using magnetic fields to manipulate the attendant physical characteristics \cite{burkov_topological_2016,klemenz_topological_2019}. How magnetic order couples to the topological quasiparticles is an interesting and multifaceted question.

Thanks to ongoing developments in theory and experiment, the catalogue of magnetic topological semimetals continues to expand. Several examples of such materials have been found among the \ch{$A$Mn{$X$}2} layered manganese pnictides ($A$ = Ca, Sr, Ba, Yb, Eu; $X$ = Bi, Sb) \cite{klemenz_topological_2019, gong_canted_2020, soh_magnetic_2019, zhang_tuning_2021, rahn_spin_2017, borisenko_time-reversal_2019, he_quasi-two-dimensional_2017, wang_two-dimensional_2012, guo_coupling_2014, zhang_effect_2022, liu_nearly_2016, wang_quantum_2011, liu_magnetic_2017, wang_magnetotransport_2016, ni_origin_2021, pal_optical_2018, kealhofer_observation_2018, qiu_observation_2019, soh_magnetic_2021, wang_quantum_2018}. In this family, Dirac or Weyl fermions are harboured by the Bi or Sb square net, which lies between magnetic Mn layers. Although many \ch{$A$Mn{$X$}2} materials  have closely related crystal structures, the Mn (and in some cases Eu) ions produce a variety of magnetic structures, which can be extended by the application of an external magnetic field. 

\begin{figure}[ht!]
	\includegraphics[width=0.5\textwidth]{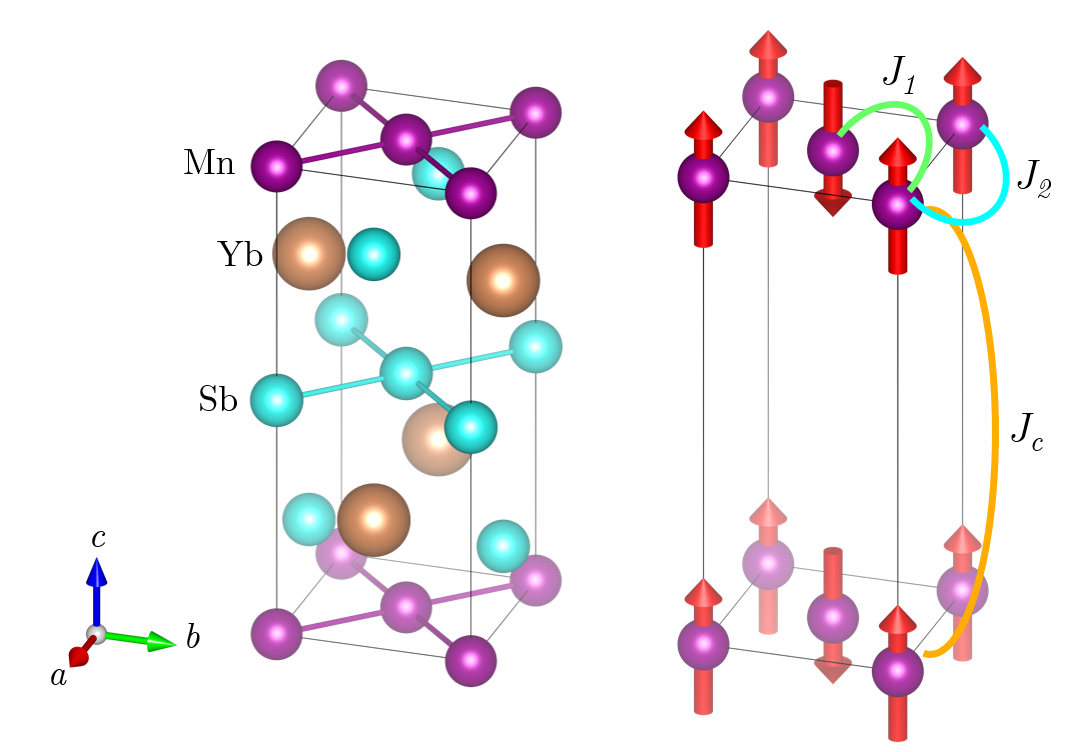}
	\caption{\label{unit_cell_fig} Left: the unit cell of \ch{YbMnSb2}. Right: the same structural unit cell showing the observed C-type antiferromagnetic ordering of the Mn spins \cite{soh_magnetic_2021}. Magnetic moments indicated by red arrows, and exchange constants labelled as $J_1$ between $ab$ plane nearest-neighbour Mn ions, $J_2$ between next-nearest neighbour Mn ions and $J_c$ along the $c$ axis. Figure prepared using VESTA \cite{momma_vesta_2011}.}
\end{figure}

In this work we shall focus on \ch{YbMnSb2}, which has been studied in detail via quantum oscillations, magnetometry, optical spectroscopy, \emph{ab initio} band structure calculations, ARPES, and single-crystal neutron diffraction \cite{kealhofer_observation_2018, qiu_observation_2019, wang_quantum_2018, soh_magnetic_2021}. \ch{YbMnSb2} is of particular note due to its superior thermoelectric properties among topological semimetals \cite{pan_thermoelectric_2021} as a result of anomalous transport in the Sb layer. The crystal structure of \ch{YbMnSb2} is described by the $P4/nmm$ space group with lattice parameters $a = b = 4.31(2)$~\AA, $c = 10.85(1)$~\AA \cite{soh_magnetic_2021}. It orders antiferromagnetically below $T_\textrm{N} \approx$~345\,K \cite{kealhofer_observation_2018, qiu_observation_2019, wang_quantum_2018, soh_magnetic_2021}. Several different magnetic structures were originally proposed based on \emph{ab initio} calculations and experimental methods that indirectly probed the magnetic structure \cite{kealhofer_observation_2018,qiu_observation_2019,wang_quantum_2018}, but our recent neutron diffraction study \cite{soh_magnetic_2021} showed conclusively that the Mn spins order in a C-type antiferromagnetic (AFM) structure with spins along the $c$ axis (the Yb atoms in \ch{YbMnSb2} are non-magnetic) -- see Fig.~\ref{unit_cell_fig}.

Here we present the results from half-polarized neutron diffraction and unpolarized neutron inelastic scattering studies of \ch{YbMnSb2}. The former was designed to study how well the magnetic moments are localised on the Mn sites or elsewhere in the unit cell, while the  latter was performed in order to measure the spin-wave spectrum. The overall aim was to determine the form and strength of the exchange interactions in \ch{YbMnSb2} and to compare them with what has been found in a number of other isostructural \ch{$A$Mn{$X$}2} compounds. The interactions along the $c$ axis are of particular interest as they are potentially mediated by  topological relativistic fermions in Sb or Bi bands near the Fermi energy. We find that the spin-wave spectrum is well described by a  Heisenberg effective spin--$\frac{1}{2}$ Hamiltonian with easy-axis anisotropy which has been used previously for \ch{$A$Mn{$X$}2}. The in-plane exchange couplings are found to be similar to those in other \ch{$A$Mn{$X$}2} compounds, but the ferromagnetic (FM) exchange along the $c$ axis is 5 to 6 times larger. The results imply that the coupling between magnetism and relativistic fermions is plays a more important role in YbMnSb$_2$ than in several related compounds containing Bi instead of Sb.

%%%%%%%%%%%%%%%%%%%%%%%%%%%%%%%%%%%%%%%%%%%%%%%%%%%%%
\section{Methods}
Single crystals in the form of platelets  of typical dimensions 4 $\times$ 4 $\times$ 0.2 mm$^3$ were grown by a flux method, as detailed in \cite{wang_quantum_2018, qiu_observation_2019}. For inelastic neutron scattering, 20 crystals with a total mass of 0.84\,g were coaligned on a 0.5\,mm thick aluminium plate with the $c$ axis perpendicular to the large face of the plate. Crystals were affixed with  hydrogen-free  CYTOP$^\textrm{TM}$ fluoropolymer \cite{rule_which_2018}. The quality of individual crystals as well as the overall coalignment of the crystals was checked using a x-ray Laue diffractometer (Photonic Science). The estimated mosaicity of the ensemble of crystals was $\sim$\ang{3} (full width at half maximum) \cite{suppl}.

Half-polarized neutron diffraction was conducted on the D3 diffractometer at the ILL \cite{lelievre-berna_ill_2005}. A single crystal of \ch{YbMnSb2} was prealigned using the ILL's neutron Laue diffractometer OrientExpress \cite{ouladdiaf_orientexpress_2006}. The crystal was oriented with the $b$ axis perpendicular to the horizontal scattering plane. A vertical-field superconducting magnet provided a magnetic field of $\mu_0 H = 9.0$\,T parallel to the $b$ axis. Polarized neutrons of wavelength $\lambda = 0.832$\,\AA were produced by a Heusler monochromator, and a cryoflipper was used to switch their spin states from parallel to antiparallel with the external magnetic field. There was no polarization analysis of the scattered neutrons (half-polarized set-up). Two erbium filters were placed in the incident beam to minimise $\lambda/2$ contamination. The sample was measured in the paramagnetic phase at a temperature of 400\,K, with any ferromagnetism manifesting solely due to the applied magnetic field. Flipping ratios (i.e. the ratio of the diffracted intensities for neutrons with spins parallel and antiparallel to the applied field) were measured at a set of structural Bragg peaks and converted to field-induced magnetic structure factors by the standard method \cite{boothroyd_principles_2020}. As the flipping ratios are all close to unity there is no ambiguity in the obtained magnetic structure factors. After averaging over symmetry-equivalent positions we obtained a total of 25 distinct magnetic structure factors at $h0l$  reflections, as well as 20 out-of-plane reflections, $hkl$ with $k = 1,2$, that were  accessible. 

 Inelastic neutron scattering was performed on the IN8 triple-axis spectrometer at the Institut Laue-Langevin (ILL) with the FLATCONE multiplexed analyser--detector system \cite{kempa_flatcone_2006}. Measurements were taken at two different crystal orientations giving access to the $(h0l)$ and $(hhl)$ sections in reciprocal space, respectively. A fixed outgoing neutron wavevector of $k_\textrm{f} = 3$\,{\AA}$^{-1}$ was selected by Bragg reflection from the silicon (Si) (111) analyser crystals built into FLATCONE, and the incident wavevector $k_\textrm{i}$ was varied to give a range of neutron energy transfers $\Delta E$ from 0 to 70\,meV. Either a double-focusing Si (111)  monochromator ($\Delta E < 40$\,meV), or a double-focusing pyrolytic graphite $(002)$ monochromator ($\Delta E \geq 40$\,meV) was used to set $k_\textrm{i}$. The FLATCONE tilt was maintained at \ang{0} throughout, and the sample was held at a temperature of 1.5\,K in a liquid helium `orange' cryostat.

%%%%%%%%%%%%%%%%%%%%%%%%%%%%%%%%%%%%%%%%%%%%%%%%%%%%%
\section{Results and discussion}
\subsection{Half-Polarized Neutron Diffraction}
\begin{figure}[h!]
	\includegraphics[width=0.45\textwidth]{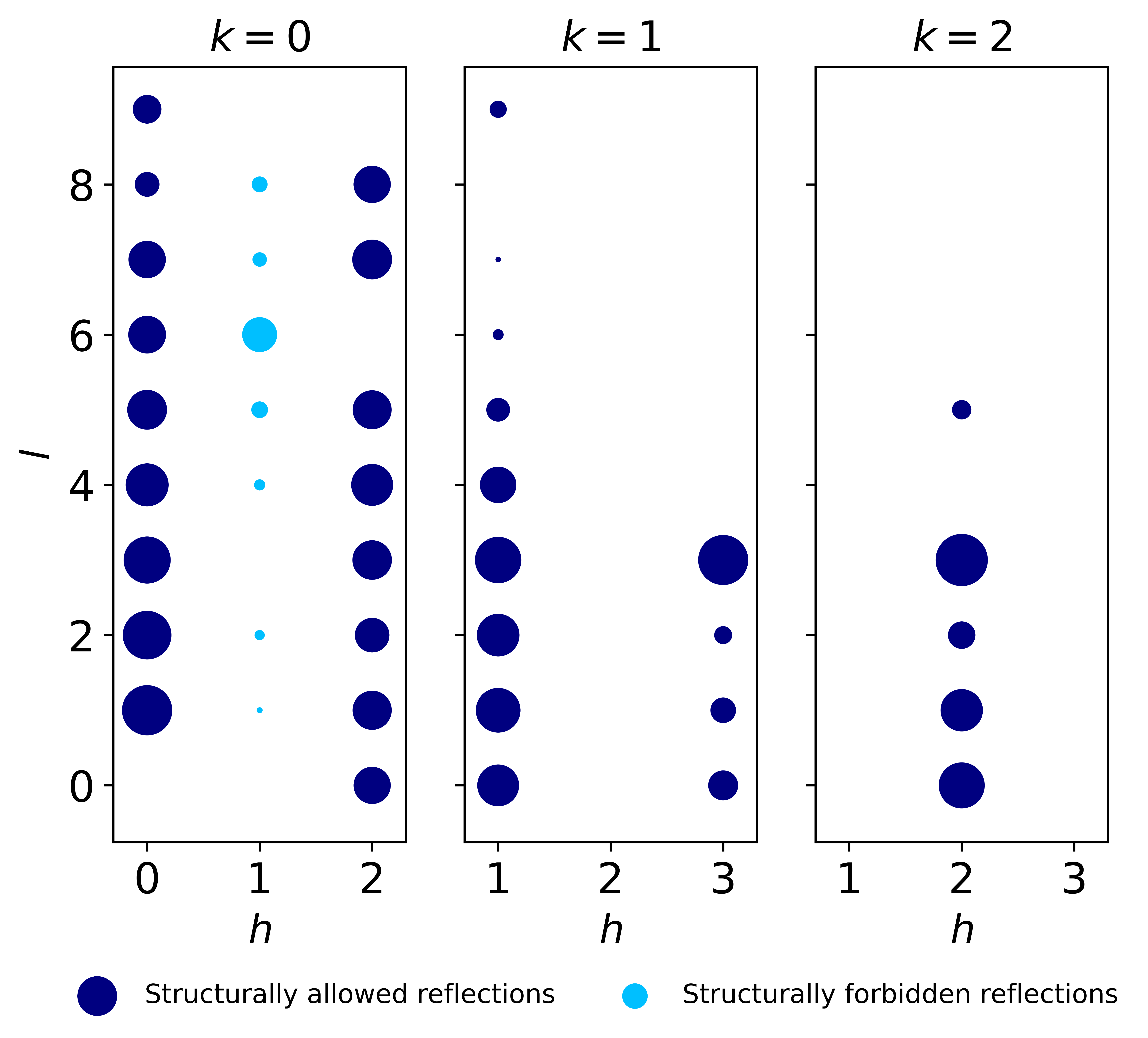}
	\caption{\label{msf_plot_fig} Magnetic structure factors calculated from measured flipping ratios. The radius of the circles is proportional to the magnetic structure factor at that $hkl$ reflection.}
\end{figure}

\begin{figure}[ht!]
	\includegraphics[width=0.45\textwidth]{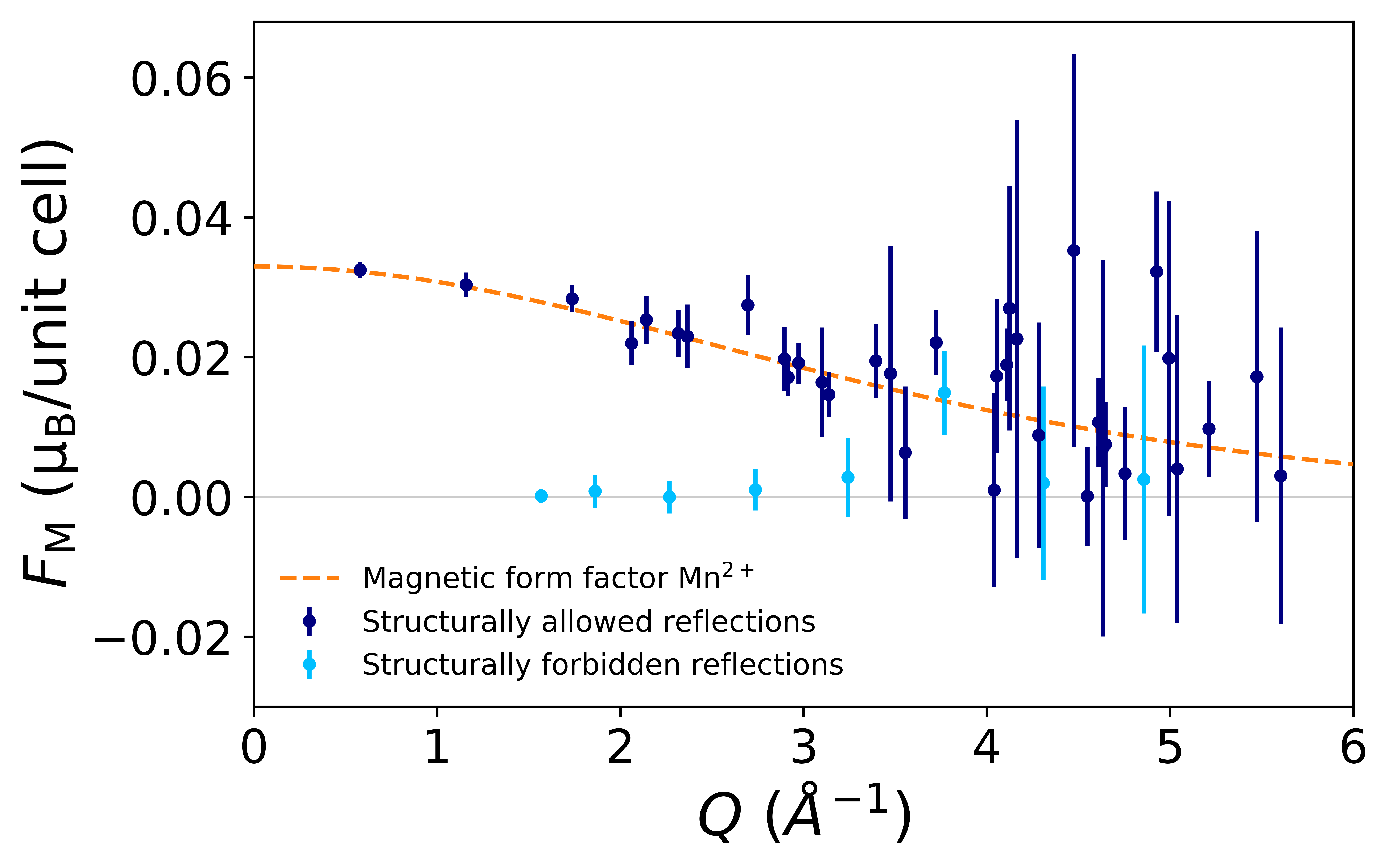}
	\caption{\label{msf_line_plot_fig} Magnetic structure factors calculated from measured flipping ratios as a function of $Q$, compared to the calculated free-ion magnetic form factor of Mn$^{2+}$.}
\end{figure}

We first describe the results from the half-polarized neutron diffraction study.  Figure \ref{msf_plot_fig} displays the positions in reciprocal space at which measurements were made, indicating the magnitude of the obtained field-induced magnetic structure factors $F_\textrm{M}$ at each position, while Fig.~\ref{msf_line_plot_fig} plots the $F_\textrm{M}$ values with their experimental uncertainties as a function of the magnitude of the scattering vector $Q = |\textbf{Q}|$.  The non-zero magnetic structure factors shown in Fig.~\ref{msf_plot_fig} obey the reflection condition $h+k = 2n$, which is consistent with the field-induced ferromagnetic phase having the same symmetry as the underlying crystal structure with space group $P4/nmm$. We also included some reflections with $h = 1$, $k = 0$ as a check. These all have zero $F_\textrm{M}$ to within the experimental error, consistent with $P4/nmm$.

\begin{figure}[ht!]
    \centering
    \includegraphics[width=0.4\textwidth]{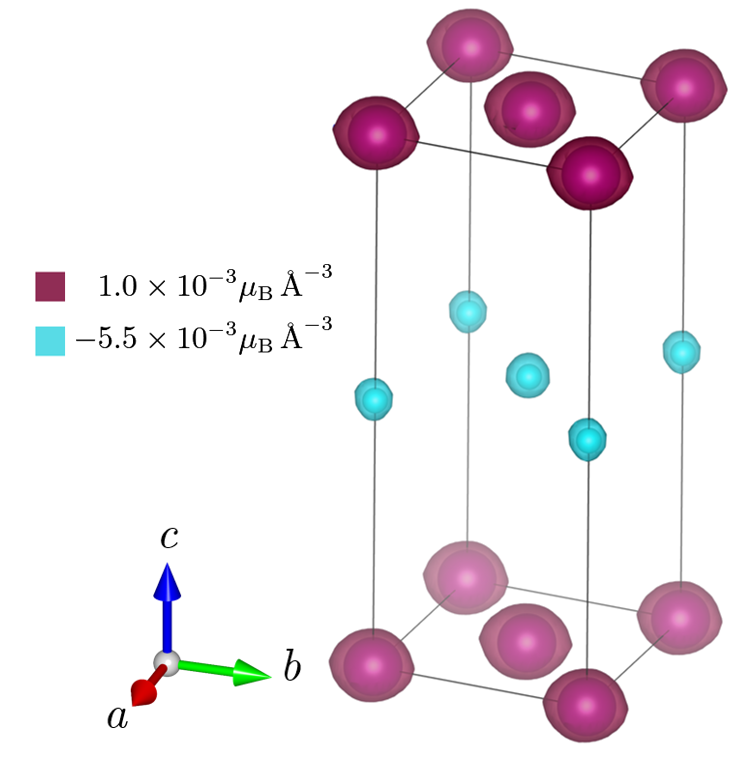}
    \caption{Magnetisation density distribution in the unit cell of \ch{YbMnSb2}, with isosurfaces showing strong concentration of magnetisation on Mn sites and a very small negative magnetisation on Sb sites in the middle of the unit cell. }
    \label{isosurface_fig}
\end{figure}

\begin{figure}[ht!]
	\includegraphics[width=0.5\textwidth]{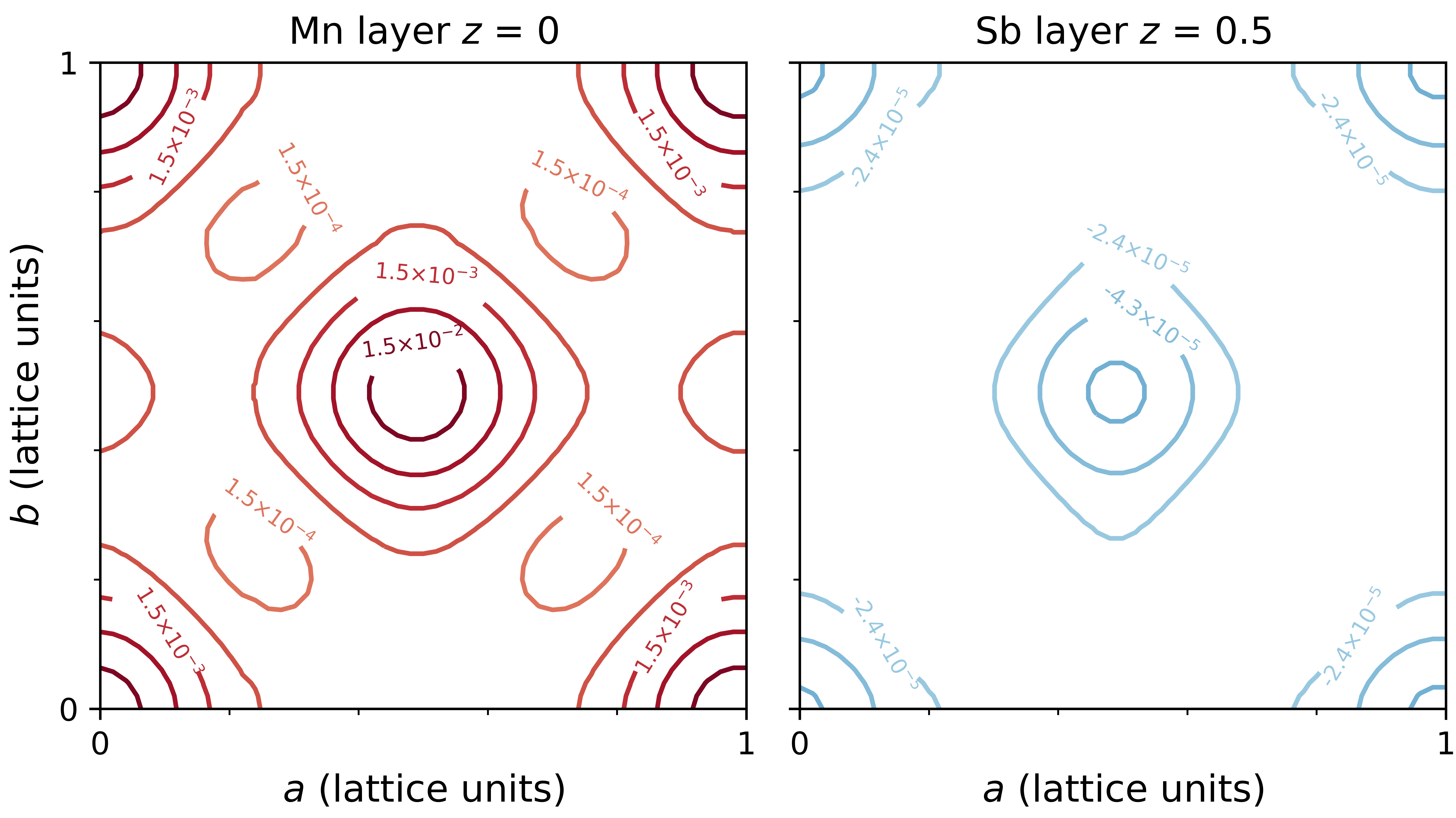}
	\caption{\label{mag_density_map_fig} Slices through the $ab$ plane showing magnetization density map reconstructed from ILL D3 data. Contour lines show magnetisation density in units of $\mu_\textrm{B}$\,\AA$^{-3}$. }
\end{figure}

A model-free reconstruction of the magnetization distribution within the unit cell (Fig.~\ref{isosurface_fig}) was obtained from the magnetic structure factors by the Bayesian maximum entropy (MaxEnt) method as implemented at the ILL, which incorporates routines from the MEMSYS subroutine library \cite{gull_memsys_1989}. Cross-sections through the reconstruction at fractional heights $z=0$ and $z = 0.5$ are presented in Fig.~\ref{mag_density_map_fig}, corresponding to the Mn layer and central Sb layer. The magnetization maps show that the majority of the magnetism induced by the external field is localized around the Mn sites and is approximately isotropic, consistent with a roughly equal population of all five Mn $3d$ orbitals as expected for the high spin  state of Mn$^{2+}$ ($3d^5, S = 5/2$). By integrating a volume around the Mn site we find the total induced magnetic moment to be 0.011(4)\,$\mu_\textrm{B}$ per Mn. A region of small negative magnetisation is observed around the Sb sites in the central layer at $z = 0.5$, with the induced moment calculated to be  $-$0.0002(6)\,$\mu_\textrm{B}$ per Sb. We used a method developed by Markvardsen to obtain these integrated moment values and their errors \cite{markvardsen_polarised_2000}. Although the MaxEnt method is designed to give a non-zero signal only when there is evidence for it in the data, the size of the uncertainty in the Sb moment is comparable with the moment itself, and so we are cautious about attributing statistical significance to this signal.  

%%%%%%%%%%%%%%%%%%%%%%%%%%%%%%%%%%%%%%%%%%%%%%%%%%%%%%%%%%%%%%%%%%%%
\subsection{Triple-Axis Spectroscopy}

\begin{figure}[ht!]
     \centering
     \includegraphics[width=\textwidth]{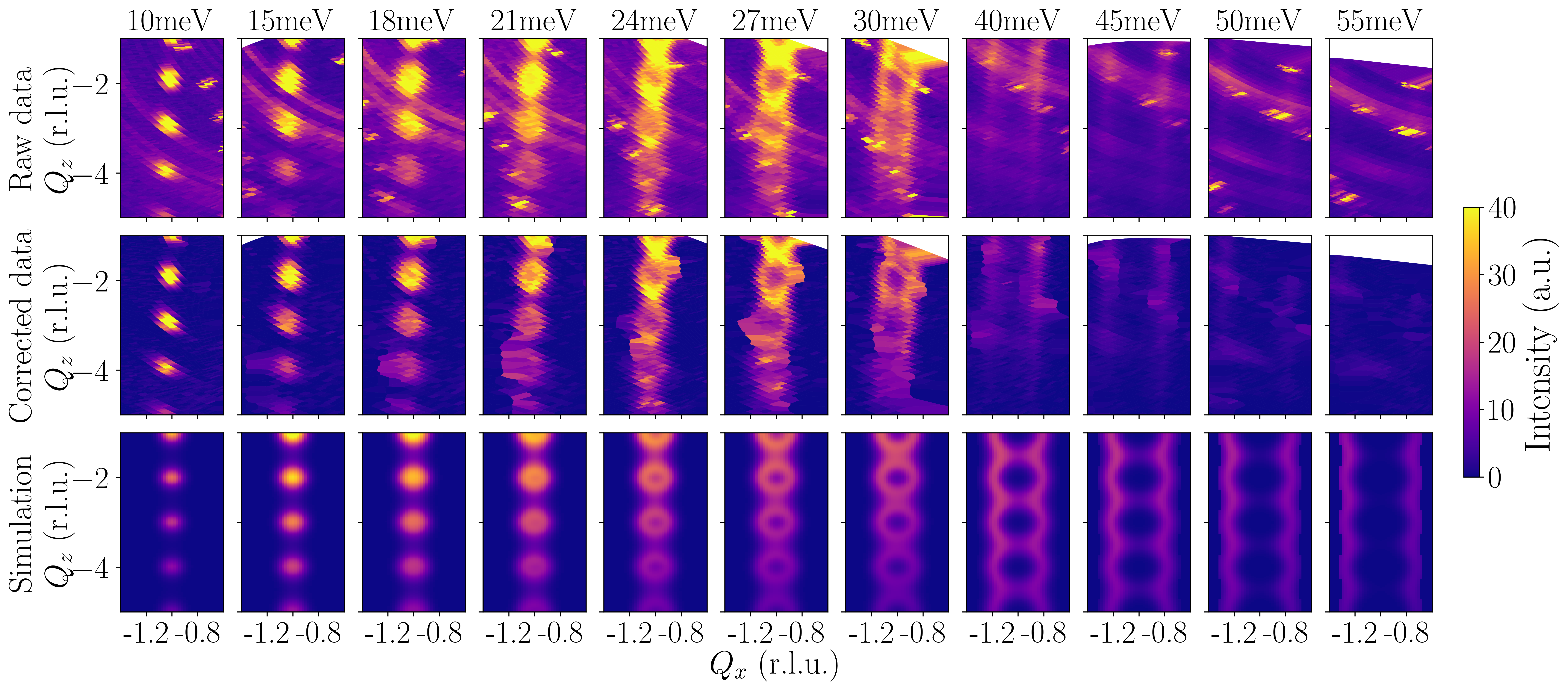}
     \caption{Spin-wave scattering and model for $h0l$ plane. Top row:  raw neutron scattering data. Middle row: Data after the corrections described in the text. Bottom: magnetic scattering simulated by linear spin-wave theory from our model. }
     \label{h0l_plane_spin_wave_fig}
 \end{figure}

\begin{figure}[ht!]
     \centering
     \includegraphics[width=\textwidth]{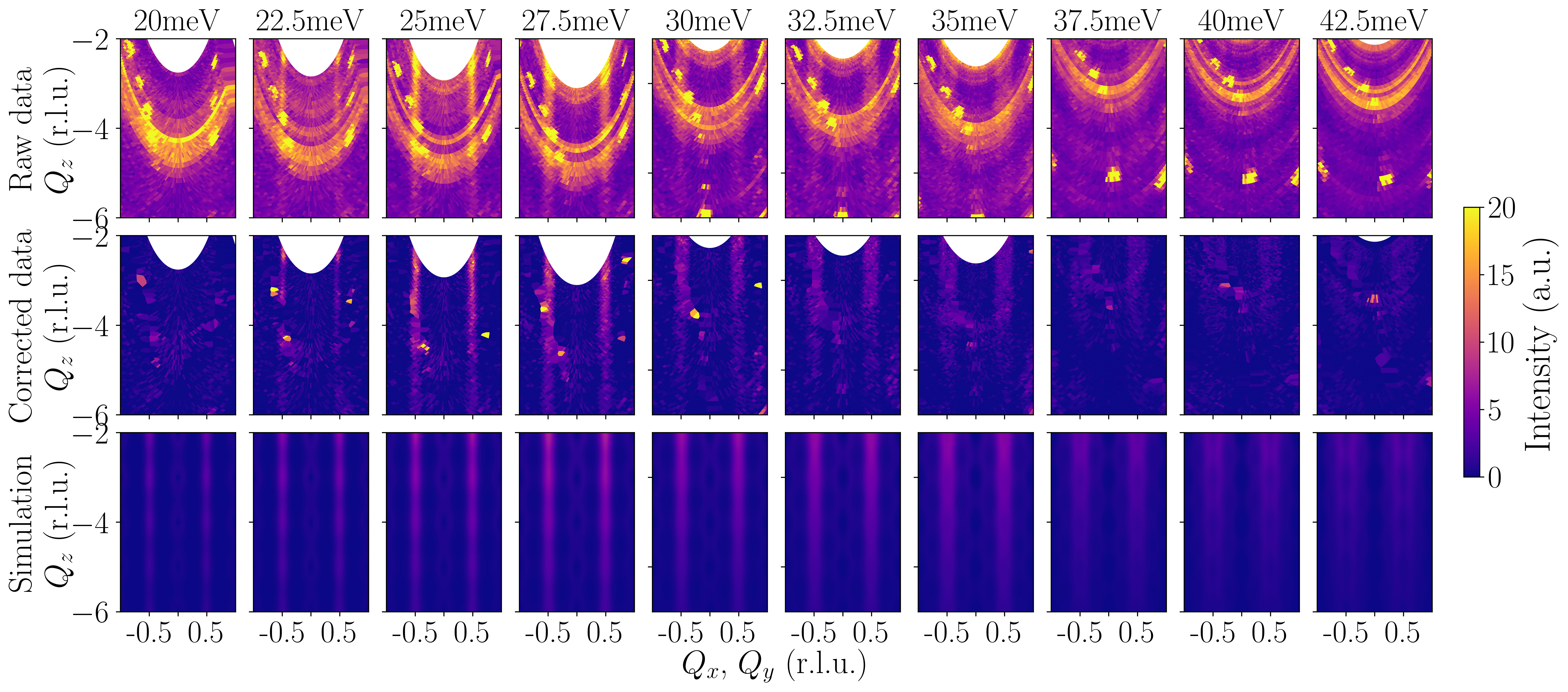}
     \caption{Spin wave data and model for $hhl$ plane. The three rows show the experimental and simulated data as described in Fig.~\ref{h0l_plane_spin_wave_fig}. }
     \label{hhl_plane_spin_wave_fig}
 \end{figure}

Moving on to the magnetic excitations, we present our neutron spectroscopy data as a series of intensity maps recorded at different energies $\Delta E$ in the $h0l$ (Fig.~\ref{h0l_plane_spin_wave_fig}) and $hhl$ (Fig.~\ref{hhl_plane_spin_wave_fig}) planes of reciprocal space. In the $h0l$ plane, the intensity is localised around the magnetic Bragg peak positions at low energies. The intensity is highest at 10\,meV, decreasing at lower energies until no signal is visible below about 5\,meV \cite{suppl}.  Above 10 meV, the intensity disperses outwards with increasing energy to form rings which eventually merge into rods of scattering running along the $(00l)$ direction. In the $hhl$ plane, rods of scattering along $(00l)$ are observed at all energies studied, with some intensity modulation along the rods at the lower energies. At higher energies the rods split into two, which then move apart. By 55 meV, the signal is almost unresolvable from the background.  As we now show, the features just described are consistent with the scattering from spin-wave excitations of antiferromagnetically ordered Mn local moments.

 %%%%%%%%%%%%%%%%%
\clearpage
\pagebreak
 %%%%%%%%%%%% spin wave data figures
\onecolumngrid

\begin{figure}[ht!]
	\includegraphics[width=0.475\textwidth]{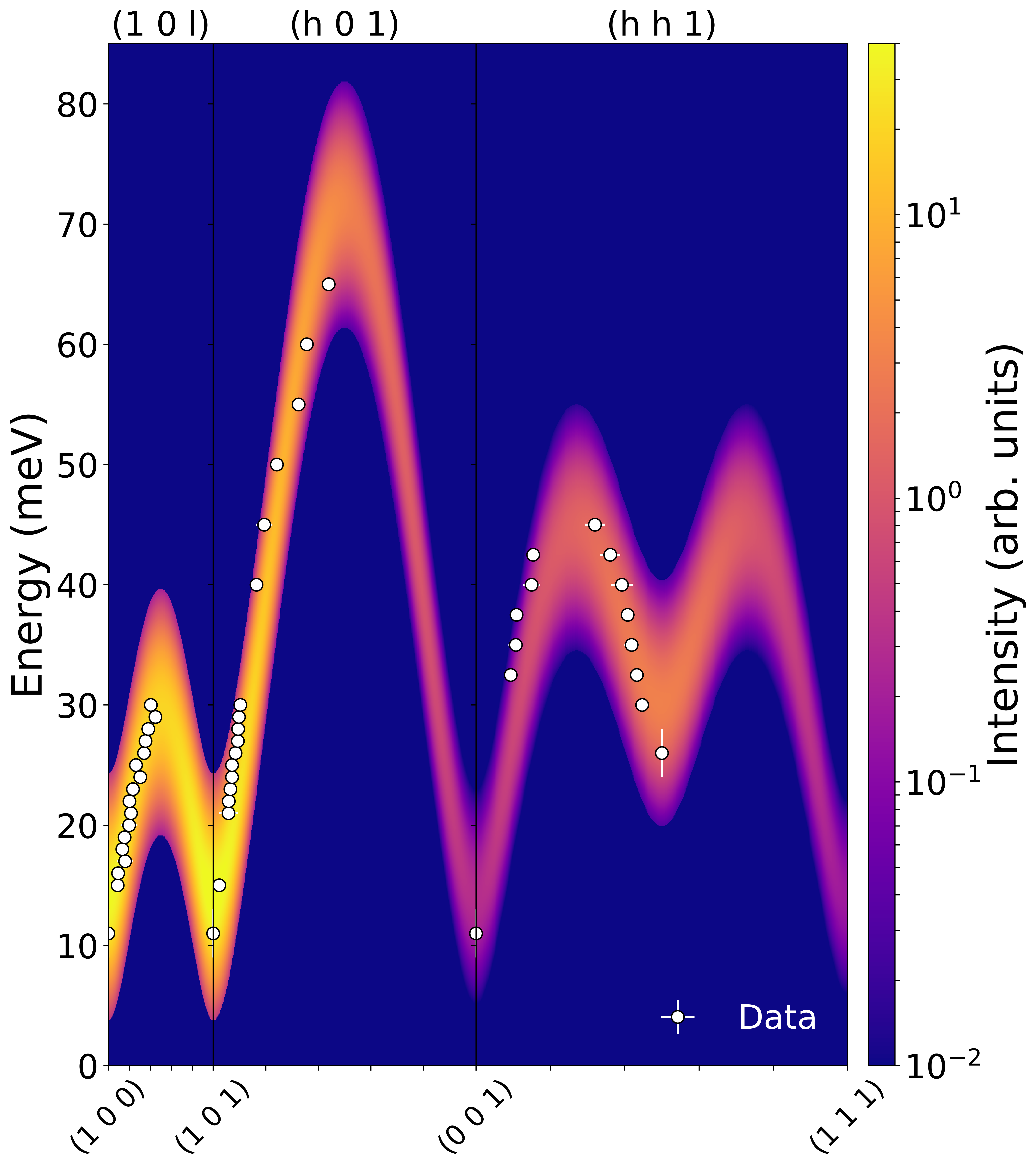}
	\caption{\label{dispersion_fig} Spin-wave dispersion of YbMnSb$_2$. The data points are shown as white circles, and the spin-wave spectrum calculated from our model is shown as an intensity map with a log color scale. The simulation includes a Gaussian energy broadening of 8 meV (FWHM).}
\end{figure}

\twocolumngrid
 %%%%%%%%%%%% end spin wave data figures
In order to analyse the  neutron inelastic scattering data \cite{suppl}, first, the \textsc{Nplot} \textsc{Matlab} scripts library \cite{steffens_nplot_2019} was used to smooth the background and manually remove spurions and powder rings.  
Second, a suite of custom \textsc{Python} programs (built using packages including those described in \cite{hunter_matplotlib_2007, harris_array_2020, virtanen_scipy_2020}) took care of visualization, interpolation, making 1D cuts through the data along high symmetry lines and intensity peak-fitting at different $\Delta E$.
The high-symmetry lines used for cuts were $\Lambda = (0, 0, l)$, $U = (h,0, \frac{1}{2})$, $\Delta = (h,0,0)$, $S = (h, h, \frac{1}{2})$ and $\Sigma = (h, h, 0)$. Additionally, the intensity maxima as a function of energy were found at the $\Gamma$ and $A$ points in order to anchor the turning points of the dispersion. Third, the extracted peak were assembled into the spin-wave dispersion along high-symmetry paths in reciprocal space shown in Fig.~\ref{dispersion_fig}. 

Following the approach used in Refs.~\cite{rahn_spin_2017,soh_magnetic_2019-1,sapkota_signatures_2020} for structurally related $A$Mn$X_2$ compounds, we modelled the magnetic interactions in YbMnSb$_2$ with a Heisenberg effective spin Hamiltonian of the form 
\begin{equation}
\mathcal H = \sum_{i,j} J_{ij} \mathbf{S}_i \cdot \mathbf{S}_j - \sum_i D \left(S_i^z \right)^2,\label{hamiltonian_eq}\end{equation}
and calculated the spin-wave spectrum using linear spin-wave theory (LSWT) for spins $\textbf{S}_i$ localized on the Mn ions. The parameters of $\mathcal H$ were obtained from a global fit to the measured dispersion curves using both custom \textsc{Python} programs (orthogonal distance regression) and the \textsc{SpinW} \textsc{Matlab} library \cite{toth_linear_2015}.  Three isotropic exchange parameters, $J_{ij} = J_1, J_2$ and $J_c$ as defined in Fig.~\ref{unit_cell_fig}, together with the single-ion anisotropy parameter $D$, which in Eq.~(\ref{hamiltonian_eq}) favours alignment of the Mn spins along the $c$ axis, were found to be sufficient to obtain a good fit to the data. 

 The calculated LSWT dispersion obtained from the best-fit parameters is shown in Fig.~\ref{dispersion_fig} along with the data points. Overall, there is a good match between the model and the data along all high symmetry directions. The fitted model indicates that the  spin-wave dispersion has a maximum energy of approximately 70\,meV and occurs at the $X$-point in the Brillouin zone, where $X = (\frac{1}{2}, 0, 0)$, and a minimum gap of about 10\,meV at the $\Gamma$-point. 
 
 A more complete comparison of the measured and calculated spin-wave spectrum can be made by inspection of Figs.~\ref{h0l_plane_spin_wave_fig} and \ref{hhl_plane_spin_wave_fig}. The bottom rows of panels show simulations of the LSWT spectrum in the same region of reciprocal space and for the same energies as probed experimentally. The magnetic form factor of Mn$^{2+}$ is included in the simulations. It can be seen that the model reproduces the distribution of measured intensity throughout energy and momentum space very well.

 The Hamiltonian parameters extracted from the LSWT fit are $SJ_1 = 28 \pm 2$\,meV, $SJ_2 = 10.4(5)$\,meV, $SJ_c = -0.73(7)$\,meV and $SD = 0.44(5)$\,meV (the spin quantum number $S$ always multiplies the Hamiltonian parameters in LSWT). Here, positive values represent antiferromagnetic coupling, while negative values imply ferromagnetic coupling. The larger uncertainty in the $SJ_1$ value may be attributed to the inherent broadening of the dispersion in the $ab$ plane, caused by the imperfect coalignment of the platelet crystals in this plane. According to the convention in Eq.~(\ref{hamiltonian_eq}), the positive value of $SD$ means that the Mn spins preferentially align along the $c$ axis, as observed experimentally \cite{soh_magnetic_2021}. The exchange constants are presented in Table~\ref{exchange_table_fig}.

For comparison, we have also included in Table~\ref{exchange_table_fig} the experimentally determined structural and exchange parameters for several other materials closely related to YbMnSb$_2$. There are a number of interesting points to note from the data. First, the $T_\textrm{N}$ of \ch{YbMnSb2} is distinctly higher than that of the other materials, hinting at a more robust magnetically ordered phase with stronger interactions. The exchange constants we have obtained reinforce this, with $SJ_1$, $SJ_2$ and $SJ_c$ all greater for \ch{YbMnSb2} than other materials in the $A$Mn$X_2$ family, especially $SJ_c$. Second, the $a$ lattice parameter of YbMnSb$_2$ is about 5\% smaller than that of the other materials, which likely explains why the $SJ_1$ and $SJ_2$ parameters are largest for YbMnSb$_2$. Third, the Mn-$X$-Mn bond angles associated with the the $J_1$ and $J_2$ superexchange paths both increase slightly, from 65.92$^{\circ}$ to 67.10$^{\circ}$ for $J_1$ and 100.60$^{\circ}$ to 102.82$^{\circ}$ for $J_2$, going from YbMnBi$_2$ to YbMnSb$_2$. These changes will tend to reduce the AFM superexchange contribution to $J_1$ and increase it for $J_2$, consistent with what is observed, although the effect is smaller than the overall increase in $J_1$ and $J_2$ associated with the decrease in $a$. 

Finally, the distance between Mn atoms in the $c$ direction is almost the same for all the compounds listed in Table~\ref{exchange_table_fig}, and yet $SJ_c$ is around five times larger in YbMnSb$_2$ than in the Bi compounds. This suggests that $J_c$ is not affected so much by differences in the exchange path geometry, but rather by the fact that the $p$ orbitals that mediate superexchange are lower in energy in Sb than in Bi. Assuming that $J_c$ is determined by a competition between FM metallic double exchange and AFM superexchange~\cite{guo_coupling_2014}, we expect a weaker AFM superexchange and hence a stronger net FM $J_c$ in Sb relative to Bi, as observed. This implies that coupling between magnetism and the relativistic fermions that occupy the Dirac-like  pnictogen bands located near the Fermi level may play a more important role in YbMnSb$_2$ than in the Bi compounds.
 
%%%%%%%%%%%%%%%%%
\clearpage\pagebreak
%%%%%%%%%%%%%%%%% 

%%%%%%%%%%%% TABLE
\onecolumngrid
\begin{table}[t!]
	\centering
 
\begin{threeparttable}
%\begin{tabular}{C{2cm}|C{1.5cm}|C{1.5cm}|m{7cm}}
\caption{Experimentally determined parameters for \ch{$A$Mn{$X$}2}. Exchange constants $J_1$, $J_2$ and $J_c$ are defined in Fig.~\ref{unit_cell_fig}, and $D$ is the easy-axis anisotropy parameter. Mn-$X$-Mn bond angles: angle 1 is with one Mn located at the corner of the unit cell and the other in the centre of $z=0$ face, and angle 2 is with the Mn located on neighbouring corners of the unit cell.  Values in parentheses are errors in the last digits.}
\label{exchange_table_fig}

\bgroup
\def\arraystretch{1.3}%  1 is the default, change whatever you need
\begin{tabular}{%
 		m{2.3cm}<{\centering}
 		m{1.5cm}<{\centering}
 		m{1.1cm}<{\centering}
  		m{1.1cm}<{\centering}
 		m{1.1cm}<{\centering}
 		m{1.4cm}<{\centering}
 		m{1.4cm}<{\centering}
 		m{1.4cm}<{\centering}
 		m{1.4cm}<{\centering}
 		m{2cm}<{\centering}
 		m{2cm}<{\centering}
 		}
\toprule
Material & Space group & $a$ (\AA)& $c$ (\AA) & $T_\textrm{N}$ (K) & $SJ_1$ (meV) & $SJ_2$ (meV)  & $SJ_c$ (meV) & $SD$ (meV) & Mn-$X$-Mn bond angle 1 ($^{\circ}$) & Mn-$X$-Mn bond angle 2 ($^{\circ}$)\\ 
\midrule
{\ch{CaMnBi2} \cite{rahn_spin_2017}} & $P4/nmm$ & 4.50 & 11.07 & 264 & 23.4(6) & 7.9(5) & -0.10(5) & 0.18(3) & 67.219(2) & 103.04(1) \\ %\hline 
{\ch{SrMnBi2} \cite{rahn_spin_2017}} & $I4/nmm$ & 4.58 & 23.14& 287 &21.3(2) & 6.39(15) & 0.11(2) & 0.31(2) & 68.047(1) & 104.614(1)\\ %\hline 
{\ch{YbMnBi2} \cite{soh_magnetic_2019-1}} & $P4/nmm$ & 4.49 & 10.86 & 290 & 22.6(5) & 7.8(5) & $-0.13(5)$ & 0.37(4) & 65.92(4) & 100.60(7) \\ 
{\ch{YbMnBi2} \cite{sapkota_signatures_2020}\tnote{a}} & $P4/nmm$ & 4.48 & 10.8 & 290 & 22.7(3) & 7.8(2) & $-0.16(3)$ & 0.43(4) &  &  \\ 
%\hline 
{\ch{YbMnSb2} (this work and \cite{soh_magnetic_2021})} &  $P4/nmm$ & 4.31(2) & 10.85(1) & 345 & 28 $\pm$ 2 & 10.4(5) & $-0.73(7)$ & 0.44(5) & 67.10(4) & 102.82(7) \\ 
\bottomrule
\end{tabular}

\egroup
\begin{tablenotes}
\item [a] {The quoted parameters are from an analysis method which is similar to that used to obtain the other parameters listed in this table. A resolution-corrected fitting method gave the following parameters: $SJ_1 = 25.9(2)$\,meV, $SJ_2 = 10.1(3)$\,meV, $SJ_c = -0.130(3)$\,meV, and $SD = 0.20(1)$\,meV.}
\end{tablenotes}
\end{threeparttable}
%\label{exchange_table_fig}
\end{table}
\twocolumngrid
%%%%%%%%%%%%%%%%%%%%%%%%%%%%%%% END TABLE

%%%%%%%%%%%%%%%%%%%%%%%%%%%%%%%%%%%%%%%%%%%%%%%%%%%%%
\section{Conclusion}

Our experiments have shown that the magnetic moments in YbMnSb$_2$ are well localized on the Mn atoms, and so a semiclassical spin-wave description of the magnetic dynamics in the antiferromagnetically ordered phase is appropriate. 

The exchange interactions determined in this work from the magnon dispersion of YbMnSb$_2$ are all larger than in several related compounds containing Bi instead of Sb which have been studied recently. The value of the interlayer coupling $J_c$ is particularly notable, being around five times larger in YbMnSb$_2$ than in the Bi compounds. Considering the different superexchange and metallic contributions to $J_c$, we argue that the coupling between Dirac fermions and local spin moments on Mn may be more prominent in YbMnSb$_2$ than in the related Bi compounds. This suggests that YbMnSb$_2$ is a promising system with which to investigate the interplay between magnetism and topological band electrons.

%%%%%%%%%%%%%%%%%%%%%%%%%%%%%%%%%%%%%%%%%%%%%%%%%%%%%
\begin{acknowledgments}
We thank Igor Mazin for insights on the exchange interactions, and all those involved in the ILL VISA project that made it possible to perform the neutron scattering experiments remotely during the coronavirus pandemic in 2020/21. Data from the neutron experiments are available from the ILL via proposal numbers 4-01-1684 (IN8 \cite{jian-rui_soh_magnon_2021}) and 5-53-305 (D3 \cite{jian-rui_soh_structural_2021}). We also benefited from the ILL crystal alignment and structural characterisation instruments CYCLOPS (experiment no. EASY-671) and Orient Express (experiment no. EASY-667). Y.F.G. and A.T.B. acknowledge support from the Oxford–ShanghaiTech collaboration project. Y.F.G. was supported by the Double First-Class Initiative Fund of ShanghaiTech University. S.M.T. was supported by a scholarship from the Rhodes Trust. J.-R.S. acknowledges support from the Singapore National Science Scholarship, Agency for Science Technology and Research, and also from the European Research Council under the European Union’s Horizon 2020 research and innovation program synergy grant (HERO, Grant No. 810451).
\end{acknowledgments}

\bibliography{main.bib}
\end{document}